\newif\iftwo 
\global\twotrue

\iftwo
\documentclass[twocolumn]{aastex631}
\else
\documentclass[manuscript,linenumbers]{aastex631}
\fi

\usepackage{wasysym} 

\newcommand{\scs}{\scriptsize}

\newcommand{\meth}{Section~\ref{sec:meth}}

\newcommand{\sm}{\mbox{$\sim$}}
\newcommand{\D}{\Delta}

\newcommand{\gtfL}{\mbox{($g_2$$-$$g_5$)}}
\newcommand{\gftL}{\mbox{($g_4$$-$$g_3$)}}
\newcommand{\sftL}{\mbox{($s_4$$-$$s_3$)}}
\newcommand{\gtf}{\mbox{$g_{25}$}}
\newcommand{\ptf}{\mbox{$P_{25}$}}
\newcommand{\sigot}{\mbox{$\sigma_{12}$}}
\newcommand{\tauot}{\mbox{$\tau_{12}$}}
\newcommand{\thtot}{\mbox{$\theta_{12}$}}
\newcommand{\vpi}{\varpi}
\newcommand{\Om}{\mbox{$\Omega$}}
\newcommand{\atant}{\mbox{atan2}}
\newcommand{\eE}{\mbox{$e_{\scs \earth}$}}
\newcommand{\eM}{\mbox{$e_{\scs \mercury}$}}
\newcommand{\x}{\times}
\newcommand{\q}{\frac}
\newcommand{\pmo}{\mbox{$^{-1}$}}

\newcommand{\beqn}{\begin{eqnarray}}
\newcommand{\eeqn}{\end{eqnarray}}
\newcommand{\lsim}{\raisebox{-.5ex}{$\stackrel{<}{\sim} \ $}}
\newcommand{\gsim}{\raisebox{-.5ex}{$\stackrel{>}{\sim} \ $}}

\newcommand{\orb}{{\tt orbitN}}
\newcommand{\zenurl}{\url{zenodo.org/records/8021040}}
\newcommand{\giturl}{\url{github.com/rezeebe/orbitN}}
\newcommand{\myurl}{\url{www2.hawaii.edu/~zeebe/Astro.html}}
\newcommand{\npurl}{\url{www.ncdc.noaa.gov/paleo/study/xxxxx}}
\newcommand{\snvurl}{\url{github.com/rezeebe/snvec}}

\def\otp{\sm{40}\%}

\renewcommand{\v}[1]{\mbox{\bm$#1$\ubm}}
\newcommand{\vb}[1]{\mbox{\boldmath$#1$}}

\shorttitle{Solar System Resonance}
\shortauthors{Zeebe \& Lantink}

\graphicspath{{./}{../eps/}{./corres/}}

\def\figgdM{\ref{fig:g.dM}}
\def\figgsFFT{\ref{fig:gsFFT}}
\def\figtheta{\ref{fig:theta}}

\begin{document}

\title{A secular solar system resonance that disrupts the dominant
cycle in Earth's orbital eccentricity ($g_2$$-$$g_5$): Implications 
for astrochronology} 


\email{zeebe@soest.hawaii.edu}

\author[0000-0003-0806-8387]{Richard E. Zeebe}
\affiliation{
School of Ocean and Earth Science and Technology, 
University of Hawaii at Manoa, 
1000 Pope Road, MSB 629, Honolulu, HI 96822, USA.
}

\author[0000-0000-0000-0000]{Margriet L. Lantink}
\affiliation{
University of Wisconsin - Madison,
     1215 W. Dayton St., Madison, WI 53706, USA.
     lantink@wisc.edu \\
     \vspace*{4ex}
     \normalsize{\rm Final revised version, in press \\
     {\it The Astronomical Journal}}
}

\begin{abstract}
The planets' gravitational interaction causes rhythmic changes in Earth's 
orbital parameters (also called Milankovi{\'c} cycles), which have powerful
applications in geology and
astrochronology. For instance, the primary astronomical eccentricity
cycle due to the secular frequency term \gtfL\
(\sm{405}~kyr in the recent past) utilized in deep-time analyses
is dominated by Venus' and Jupiter's orbits, aka long eccentricity cycle. The
widely accepted and long-held view is that \gtfL\ was practically stable in the
past and may hence be used as a ``metronome''
to reconstruct accurate ages and chronologies. However, using state-of-the-art
integrations of the solar system, we show here that \gtfL\ can become unstable over
long time scales, without major changes in, or destabilization of, planetary orbits.
The \gtfL\ disruption is due to the secular resonance 
$\sigot = (g_1 - g_2) + (s_1 - s_2)$,
a major contributor to solar system chaos. We demonstrate that
entering/exiting the \sigot\ resonance is a common phenomenon on long time scales,
occurring in \sm{40}\% of our solutions. During \sigot-resonance 
episodes, \gtfL\ is very weak or absent and Earth's orbital eccentricity and 
climate-forcing spectrum are unrecognizable compared to the recent past. Our 
results have fundamental implications for geology and astrochronology,
as well as climate forcing because
the paradigm that the longest Milankovi{\'c} cycle dominates Earth's astronomical
forcing, is stable, and has a period of \sm{405}~kyr requires revision.
\end{abstract}



\section{Introduction} \label{sec:intro}

\begin{figure*}[t]
\vspace*{-35ex} \hspace*{-05ex}
\includegraphics[scale=0.65]{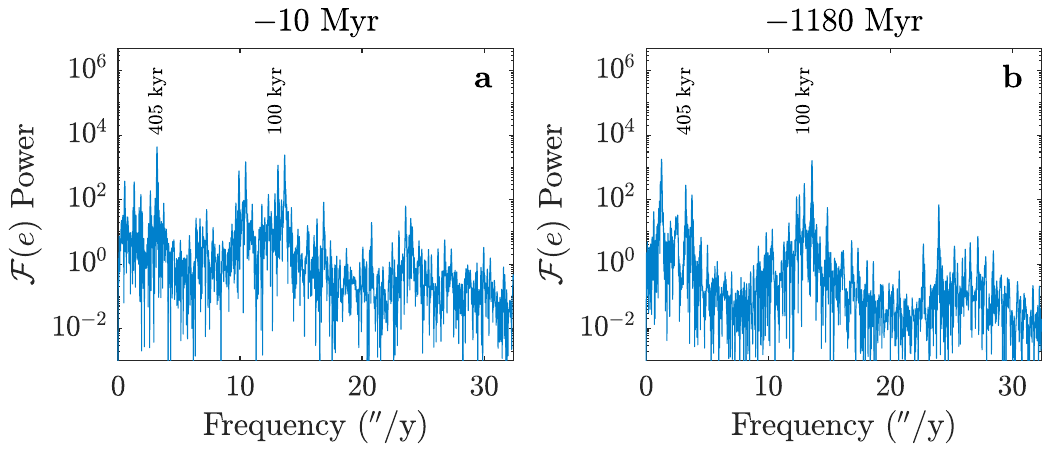}\hspace*{-13ex}
\includegraphics[scale=0.65]{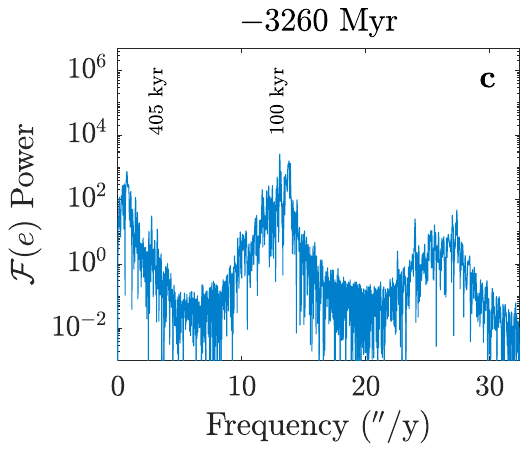}

\vspace*{-40ex}
\caption{\scs
Fast Fourier transform (FFT~$= {\cal F}(e)$) 
over 20-Myr intervals of Earth's orbital eccentricity. 
Frequencies in arcsec~y\pmo\ = ''~y\pmo.
(a) Standard spectrum in the recent past centered at
$t = -10$~Myr (nearly identical in all solutions). 
Note the dominant \gtfL\ 405-kyr LEC.
(b) and (c) Spectra of solutions R06 (Run 06) and R45 during 
\sigot-resonance episodes (see Section~\ref{sec:res})
centered at $t = -1180$~Myr and 
$t = -3260$~Myr, respectively.
Note the unrecognizable spectrum pattern
in (c) compared to (a) and the reduced/absence 
of power around the LEC frequency of
\sm{3.2}''~y\pmo\ (\sm{405}~kyr) in (b) and (c). 
The peaks around 10 to 13''~y\pmo\ are due to the 
short eccentricity cycle.
\label{fig:EccFFT}
}
\end{figure*}

Laying the foundations of chaos theory, Henri Poincar{\'e} wrote:
``It may happen that small differences in the initial conditions 
produce very great ones in the final phenomena. A small error in 
the former will produce an enormous error in the latter. Prediction 
becomes impossible $\ldots$'' \citep{poincare1914}. In reference to
the solar system, the sensitivity to initial conditions is indeed
a key feature of 
the large-scale dynamical chaos in the system, which has been 
confirmed numerically \citep{sussman88,laskar89,ito02,morbidelli02,
varadi03,batygin08,zeebe15apje,brownrein20,abbot23}. Dynamical chaos 
affects the secular frequencies $g_i$ and $s_i$ (see 
Appendix~\ref{sec:gs}), where the terms \gftL\ and \sftL, for instance,
show chaotic behavior already on a 50-Myr time scale. As a 
result, astronomical solutions diverge around $t = \pm50$~Myr,
which fundamentally prevents identifying a unique solution on 
time scales $\gsim$10$^8$~y \citep{laskar04Natb,zeebe17aj,zeebelourens19}. 
The chaos therefore not only severely limits
our understanding and ability to reconstruct and predict the 
solar system's history and long-term future, it also imposes strict 
limits on geological and astrochronological applications
such as developing a fully calibrated astronomical time scale 
beyond \sm{50}~Ma \citep[for recent efforts, see][]
{zeebelourens19,zeebelourens22epsl}. 
In contrast to these limitations (largely due to unstable terms 
such as \gftL\ and \sftL), another frequency term appears
more promising as it shows more 
stable behavior. For example, it has hitherto been assumed that
\gtfL\ was practically stable in the past and has been suggested
for use as a ``metronome'' in deep-time geological applications, 
i.e., far exceeding 50~Ma \citep{laskar04Natb,kent18,
spalding18,meyers18,montenari18,lantink19,devleesch24}.
The \gtfL\ cycle, which is the dominant term in Earth's orbital
eccentricity in the recent past (\sm{405}~kyr, 
see Fig.~\ref{fig:EccFFT}a) may thus have been regarded as an 
island of stability in a sea of chaos. However, we show in this 
contribution that also \gtfL\ can become unstable over long time 
scales.

Solar system dynamics affect Earth's orbital evolution, as well as 
Earth's climate, which is paced by astronomical cycles
on time scales $\gsim$10~kyr. The cycles include precession and 
obliquity cycles of
Earth's spin axis (\sm{20} and 40~kyr in the recent past), 
and the short and long eccentricity cycles (\sm{100} and 
405~kyr, see Figs.~\ref{fig:EccFFT}a and~\ref{fig:ecc}) 
\citep{milanko41Natb,
montenari18}. Orbital eccentricity is controlled by the solar 
system's orbital dynamics
and is the focus 
of this study. The primary tuning target used in astrochronology and 
cyclostratigraphy for deep-time stratigraphic age models 
is the long eccentricity cycle (LEC)
because it is widely assumed to be
stable in the past (see above). 
Dynamically, Earth's orbital eccentricity and inclination
cycles originate from combinations 
of the solar system's fundamental frequencies, called $g$- and 
$s$-frequencies (or modes), loosely related to the apsidal and 
nodal precession of the planetary orbits (Fig.~\ref{fig:gsill}). 
The LEC is dominated by Venus' and Jupiter's orbits, viz., 
($g_2-g_5)$, or \gtf\ for short, and represents the strongest
cycle in Earth's eccentricity spectrum in the recent past
(see Figs.~\ref{fig:EccFFT} and~\ref{fig:ecc}). 
Assuming a stable LEC may appear 
plausible because Jupiter (dominating $g_5$)
is the most massive planet
and less susceptible to perturbations. Astronomical 
computations have confirmed $g_5$'s stability
and hitherto did not indicate instabilities in \gtf\ 
\citep{berger84nato,quinn91,varadi03,
laskar04Natb,zeebe17aj,spalding18,zeebelourens22epsl}.
However, compared to $g_5$, $g_2$'s long-term stability is 
less certain but has been overlooked so far.
The long-term stability of \gtf\ and the LEC is critical 
for, e.g., climate forcing/insolation,
constructing accurate age models and chronologies,
expanding the evidence for the astronomical theory of 
climate into the more distant past, extending 
the astronomically calibrated geological time scale into 
deep time, and more (see discussion).
Below we show that the LEC can become unstable
over long-time scales owing to $g_2$, without major changes 
in, or destabilization of, planetary orbits. Orbital
destabilization is a known, separate dynamical phenomenon 
relevant to the future, see below.

For the present study,
we performed state-of-the-art solar system integrations,
including the eight planets and Pluto, a lunar contribution, 
general relativity, the solar quadrupole moment, and solar mass loss
(see \meth) \citep{zeebe17aj,zeebelourens19,zeebe22aj,zeebe23aj}.
Initial conditions at time $t_0$ were taken from the
latest JPL ephemeris DE441 \citep{park21de} and the equations
of motion were numerically integrated to $t = -3.5$~Gyr
(beyond $-3.5$~Gyr parameters such as the lunar 
distance have large uncertainties, see \meth).
Owing to solar system chaos, the solutions diverge around $t = -50$~Myr,
which prevents identifying a unique solution on time scales 
$\gsim$10$^8$~y (see above). Hence we present
results from long-term ensemble integrations to explore the 
possible solution/phase space of the system (see \meth).
Importantly, because of the chaos, each \sm{$10^8$}~y interval 
of the integrations represents a snapshot of the system's 
general/possible
behavior that is largely independent of the actual numerical time of 
a particular solution (provided here that $t < -\tauot$, where 
\tauot\ is of order $10^8$ to $10^9$~y, see below).
In other words, a numerical solution's behavior around,
say, $t = -1.5$~Gyr may represent the actual solar system 
around $t = -600$~Myr and so on. 

\section{Methods \label{sec:meth}}

\subsection{Solar System Integrations}

\begin{figure*}[t]
\vspace*{-40ex} \hspace*{-00ex}
\includegraphics[scale=0.8]{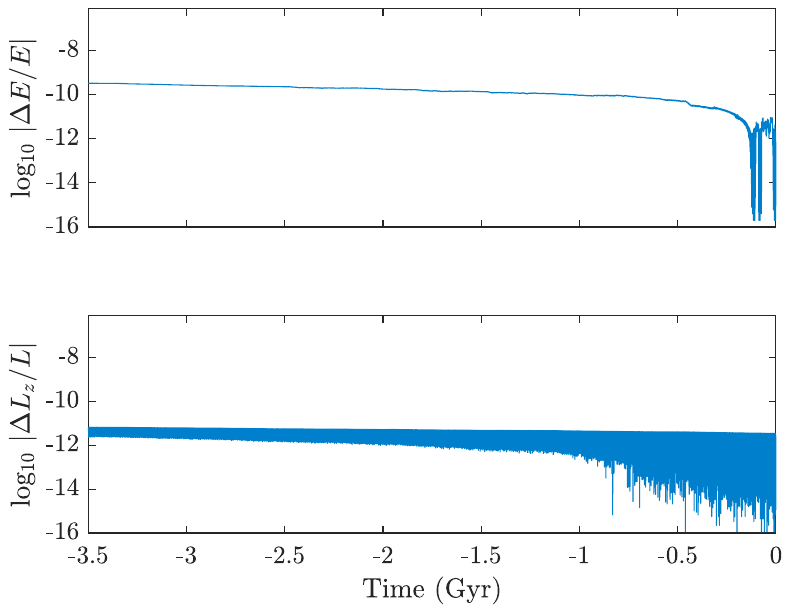}

\vspace*{-40ex}
\caption{\scs
Relative errors in total energy ($E$) and angular 
momentum ($L_z$) of 3.5-Gyr test integrations (solar mass = $const.$).
$|\Delta E/E| = |E(t) - E(0)/E(0)|$,
$|\Delta L_z/L| = |L_z(t) - L_z(0)/L(0)|$.
Due to $J_2$, only $L_z$ is conserved, not the horizontal 
components.
\label{fig:dEL}
}
\end{figure*}

Solar system integrations were performed following
our earlier work 
\citep{zeebe17aj,zeebelourens19,zeebe22aj} 
with our integrator 
package {\tt orbitN (v1.0)} \citep{zeebe23aj},
using a symplectic integrator
and Jacobi coordinates \citep{wisdom91,zeebe15apjA}. 
The open source code is available at \zenurl\ and \giturl.
The methods used here and our 
integrator package have been extensively tested and compared
against other studies \citep{zeebe17aj,zeebelourens19,zeebe22aj,
zeebe23aj}.
For the present study, we also included simulations with 
an independent integrator package (HNBody) \citep{rauch02}
and found the same dynamical behavior.
All simulations include contributions from general 
relativity, available in \orb\ as Post-Newtonian effects 
due to the dominant mass. 
The Earth-Moon system was modeled as a gravitational 
quadrupole \citep{quinn91,varadi03,zeebe17aj,zeebe23aj}.
Initial conditions for the positions and velocities of 
the planets and Pluto were generated from the 
latest JPL ephemeris DE441 \citep{park21de}
using the SPICE toolkit for Matlab. 
Coordinates were obtained at JD2451545.0 
in the ECLIPJ2000 reference frame and subsequently 
rotated to account for the solar quadrupole moment 
($J_2$) alignment with the solar rotation axis 
\citep{zeebe17aj}. Solar mass loss was included
using $\dot M/M = -7 \x 10^{-14}$~y\pmo\
\citep[e.g.,][]{quinn91}. As solar mass loss causes a 
secular drift in total energy, we added test runs 
with $M = const.$ to check the integrator's numerical 
accuracy. Total energy and angular momentum errors were 
small throughout the 3.5-Gyr integrations (relative errors: 
$\lsim$$6 \x 10^{-10}$ and $\lsim$$7 \x 10^{-12}$, 
see Fig.~\ref{fig:dEL}).
Our default numerical timestep ($|\D t| = 4$~days) is close to the 
previously estimated value of 3.59~days to sufficiently 
resolve Mercury's perihelion \citep{wisdom15,hernandez22,abbot23}.
In additional eight simulations, we tested $|\D t| = 2.15625$~days
(adequate to $\eM \ \lsim 0.4$) and found no difference in terms of 
\sigot-resonances (see below), which occurred in 3/8 solutions.

\subsection{Ensemble Integrations}
We performed ensemble integrations of the solar system
with a total of $N = 64$ members.
Note that a larger $N$ is unnecessary for the 
current problem, which samples a common phenomenon 
(\otp\ of solutions), not a rare event, which
requires large $N$ \citep{abbot23}.
Different solutions were obtained by offsetting
Earth's initial position by a small distance 
(largest offset $\D x_0 \simeq 1\x 10^{-12}$~au), which is 
within observational uncertainties \citep{zeebe15apje,
zeebe17aj}. The different $\D x_0$ lead to complete
randomization of solutions on a time scale of \sm{50}~Myr
due to solar system chaos. We also tested different 
histories of the Earth-Moon distance ($R$), which has 
little effect on our results (see Section~\ref{sec:al}). 
Because of
the large uncertainties in $R$ prior to \sm{$3$}~Ga,
we restrict our integrations to $t = -3.5$~Gyr.
Our solutions are available at \npurl\ and \myurl.

\subsection{Past Earth-Moon distance \label{sec:al}} 

\begin{figure*}[t]
\vspace*{-35ex} \hspace*{+05ex}
\includegraphics[scale=0.7]{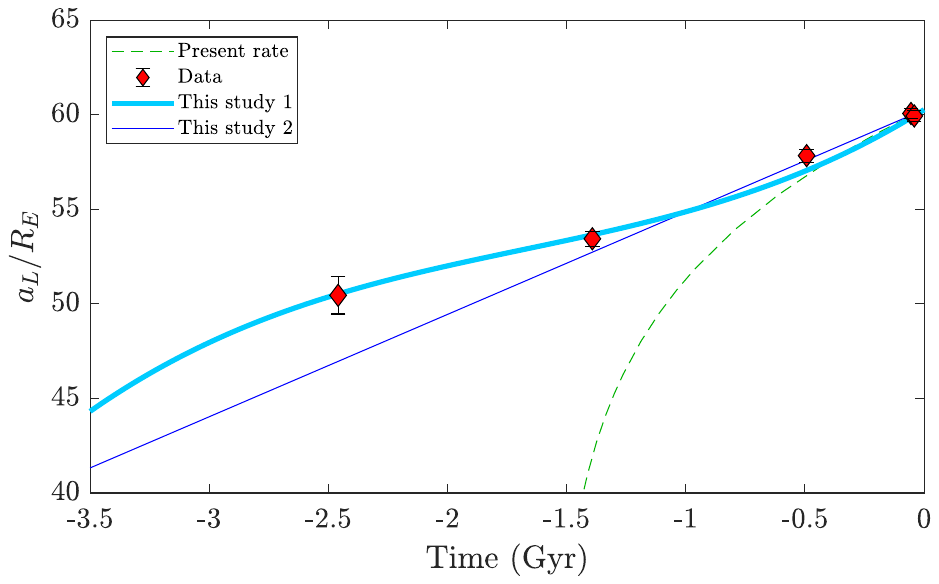}

\vspace*{-38ex}
\caption{\scs
Past Earth-Moon distance ($R$) in units of Earth radii ($R_E$).
Green dashed line: Integration of precession equations starting
at present rate $dR_0/dt$ (see text), yielding (well-known)
unrealistic past $R/R_E$. Red diamonds: Observational estimates 
based on robust data sets from cyclostratigraphic studies.
Blue and cyan lines: Used in the present study. Blue:
linear extrapolation into the past starting with $dR_0/dt$
close to the present rate. Cyan: 3rd-order polynomial fit 
to observations. Using $R/R_E$ based on the blue and cyan lines
made essentially no difference in our computations; both approaches 
yielded solutions including \sigot-resonance intervals at a similar 
frequency (see Section~\ref{sec:res}).
\label{fig:rr0}
}
\end{figure*}

Our integrations
included a lunar contribution, i.e., a gravitational quadrupole 
model of the Earth-Moon system \citep{quinn91,varadi03,zeebe17aj,zeebe23aj}.
In the present context, the lunar contribution has a relatively 
small effect on the overall dynamics, yet the integration
requires the Earth-Moon 
distance ($R$) as parameter at a given time in the past. We tested
two approaches, both avoiding the known problem of
unrealistically small $R$ at $-3.5$~Gyr
(see Fig.~\ref{fig:rr0}). ($i$) A linear extrapolation 
of $R$ into the past starting with $dR_0/dt$ close to the present 
rate and ($ii$) a 3rd-order polynomial fit to observations. 
The two approaches made essentially no difference in our computations
and both yielded solutions including \sigot-resonance intervals
at a similar frequency (see below). For the observational constraints 
on $R$, we selected robust data sets based 
on the reconstruction of Earth's axial precession frequency obtained 
by cyclostratigraphic studies \citep{meyers18,lantink22,
soerensen20,devleesch23}
(see Fig.~\ref{fig:rr0}).
The classical integration of precession equations starting
at the present rate $dR_0/dt$ (see Fig.~\ref{fig:rr0}, green dashed 
line) follows \citet{macdonald64,goldreich66,touma94}.

\subsection{Time series analysis of astronomical solutions}

The solar system's fundamental $g$- and $s$-frequencies 
were determined from the output of our numerical integrations
using fast Fourier transform (FFT) over consecutive 20-Myr 
intervals. For the spectral analysis (see Figs.~\figgdM\ 
and~\figgsFFT), we used Earth's orbital elements and 
the classic variables:
\beqn
h =  e \sin(\vpi)         \quad & ; & \quad
k =  e \cos(\vpi)         \label{eqn:hk} \\
p = \sin (I/2) \ \sin \Om \quad & ; & \quad
q = \sin (I/2) \ \cos \Om \label{eqn:pq} \ ,
\eeqn
where $e$, $I$, $\vpi$, and $\Om$ are eccentricity, 
inclination, longitude of perihelion, and longitude
of ascending node, respectively.
The spectra for Earth's $k$ and $q$, for example, show
strong peaks at nearly all $g$- and $s$-frequencies,
respectively (see Fig.~\figgsFFT).
The $g$- and $s$-modes are loosely related 
to the apsidal and nodal precession of the planetary orbits 
(see Fig.~\ref{fig:gsill}).
However, there is generally no simple one-to-one relationship 
between a single mode and a single planet, 
particularly for the inner planets. The system's motion is 
a superposition of all modes, although for the outer planets, 
some modes are dominated by a single planet. 

\subsection{Resonant angle \label{sec:rang}}

The resonant angle \thtot\ associated with the \sigot\ resonance
(see Eq.~(\ref{eqn:sig})) was determined following
\citet{lithwick11}. The method is numerically
efficient and easy to implement. Consider Eqs.~(\ref{eqn:hk}) 
and~(\ref{eqn:pq}), and use 
$\sin(I/2) \simeq I/2$ (applicable to small $I$, as in our solutions). 
The variable pairs $(h,k)$ and $2(p,q)$ can then be combined 
into two complex variables for each planet ($\hat{\imath}
= \sqrt{-1}$):
\beqn
z_k     & = & e_k \exp(\hat{\imath} \ \vpi_k) \\
\zeta_k & = & I_k \exp(\hat{\imath} \ \Om_k ) \ ,
\eeqn
where index $k = M, V, E, \ldots, N$ refers to the planets.
The $(z_k, \zeta_k)$ for $k = M, V$, for instance, were determined 
from Mercury's and Venus' computed orbital elements. Next, we
applied a simple bandpass filter (rectangular window) centered on the 
fundamental frequencies $g_i$ and $s_i$ of interest (index $i$).
For example, for $i = 1,2$, the passed frequency range was set to 
$[g_1 \ g_2] \pm 10\%$ and $[s_1 \ s_2]^{+20\%}_{-10\%}$. The
filtered (complex) quantities $(z^*_i, \zeta^*_i)$ then represent
variables in which the magnitudes $|z^*_i|, |\zeta^*_i|$
are related to the planets, $|z^*_i| \simeq e_k$ and 
$|\zeta^*_i| \simeq I_k$ (see Fig.~\figtheta\ 
and \citet{lithwick11} for details). 
The phase angles $\vpi^*_i$ and $\Om^*_i$ 
are related to the fundamental modes, 
$\vpi^*_i = \atant \{ \Im(z^*_i), \Re(z^*_i) \}$ and
$\Om^*_i  = \atant \{ \Im(\zeta^*_i), \Re(\zeta^*_i) \}$,
where $\Im$ and $\Re$ denote the imaginary and real part
of a complex number.
Finally, the resonant angle $\thtot$ associated with
\sigot\ (see Eq.~(\ref{eqn:sig})) is calculated as:
\beqn
\thtot = (\vpi^*_1 - \vpi^*_2) + (\Om^*_1 - \Om^*_2) \ .
\label{eqn:tht}
\eeqn

\section{Results \label{sec:res}}

\subsection{Secular frequencies and eccentricity}

Contrary to expectations, we found in \otp\ of the 
solutions that \gtf\ was not stable at a period $\ptf \simeq 405$~kyr 
but shifted abruptly due to shifts in $g_2$
(Fig.~\ref{fig:g.dM}). Importantly, in those cases the $g_2$ 
spectral peak usually split into two peaks at 
significantly reduced power (Fig.~\ref{fig:gsFFT}),
resulting in a very weak or absent LEC (Fig.~\ref{fig:ecc}).
Note that for, e.g., geological applications, the weak/absent
LEC is crucial, not the actual value of the \ptf\ shift 
(Fig.~\ref{fig:T25}), which is 
immaterial because it would be unidentifiable in a
stratigraphic record owing to the low \gtf\ power.
Time series analysis of the solutions
(see \meth) revealed that the weak LEC intervals 
are associated with a secular resonance between the
$g$- and $s$-modes dominated by Mercury and Venus
($|g_1 - g_2| \simeq |s_1 - s_2|$), dubbed \sigot:
\beqn
\sigot = (g_1 - g_2) + (s_1 - s_2) \ .
\label{eqn:sig}
\eeqn

\iftwo \def\sc{1.1} \def\hs{-115ex} 
\else  \def\sc{1.1} \def\hs{-095ex} 
\fi
\begin{figure*}[p]
\vspace*{-30ex} \hspace*{-45ex}
\mbox{
\includegraphics[scale=1.1]{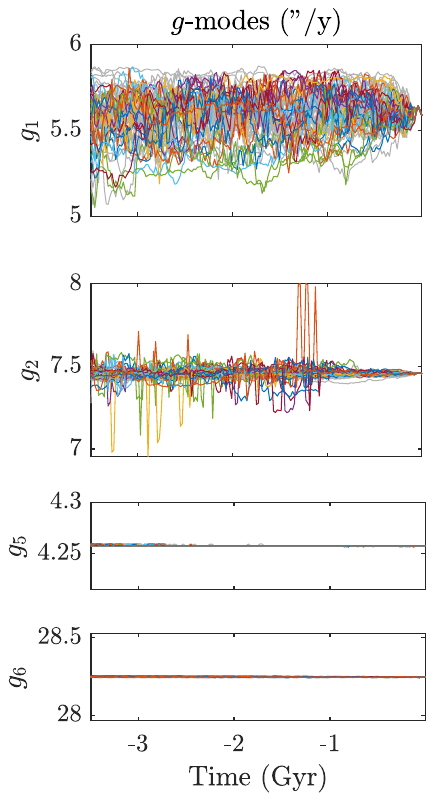}
\vspace*{+00ex} \hspace*{\hs}
\includegraphics[scale=1.1]{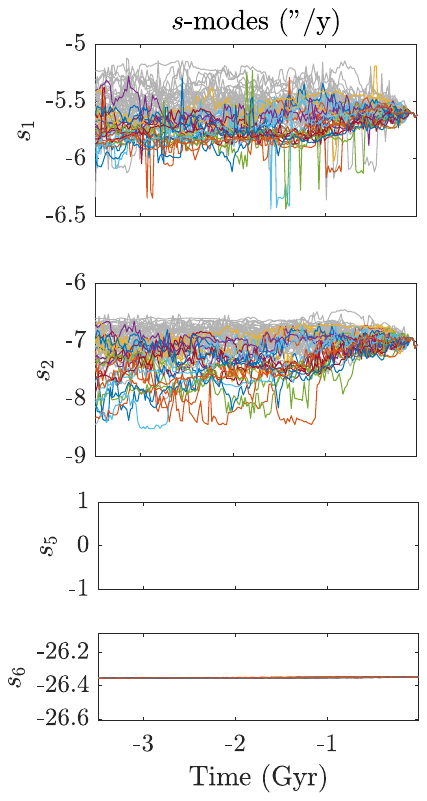}
}

\vspace*{-60ex}
\caption{\scs
Evolution of fundamental solar system frequencies. The $g$- and 
$s$-frequencies (in arcsec~y\pmo\ = ''~y\pmo)
were determined from our solar system integrations using 
fast Fourier transform (FFT) over consecutive 20-Myr 
intervals and Earth's $k$ and $q$ variables (see \meth). 
The $g$- and $s$-modes are loosely related 
to the apsidal and nodal precession of the planetary orbits 
(see Fig.~\ref{fig:gsill}).
Solutions including \sigot-resonance intervals (\otp)
are highlighted in color, the remaining solutions are displayed
in gray. The frequencies $g_1$, $s_1$, and $s_2$ drift most strongly 
over time owing to chaotic diffusion. In addition, $g_2$ shows
large and rapid shifts (spikes) at specific times when the 
spectral $g_2$ peak splits into two peaks at significantly reduced
power during \sigot-resonance episodes (see Fig.~\ref{fig:gsFFT}).
Alternating maximum power
between the two peaks then causes the spikes in $g_2$. As a result, 
$\gtf = (g_2-g_5)$ is unstable and weak/absent during \sigot-resonance
intervals (see Fig.~\ref{fig:T25}). 
$g_5$, $g_6$, and $s_6$ (dominated by Jupiter and Saturn)
are practically stable over 3.5~Gyr ($s_5$ is zero
due to conservation of total angular momentum/existence
of an invariable plane, see Fig.~\ref{fig:gsill}).
\label{fig:g.dM}
}
\end{figure*}

\iftwo \def\sc{0.8} \def\vs{-115ex} 
\else  \def\sc{0.8} \def\vs{-095ex} 
\fi
\def\hsp{+00ex}
\begin{figure*}[t]
\vspace*{-45ex} \hspace*{\hsp}
\includegraphics[scale=0.8]{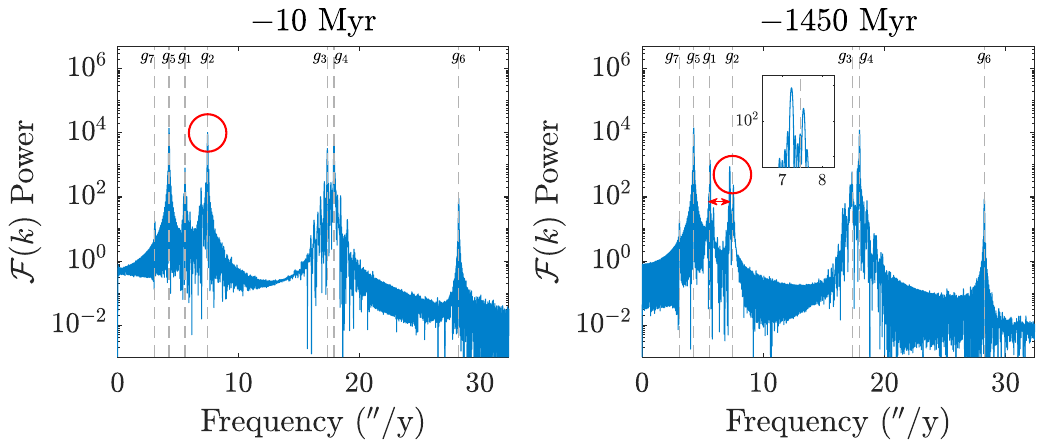}

\vspace*{\vs} \hspace*{\hsp}
\includegraphics[scale=0.8]{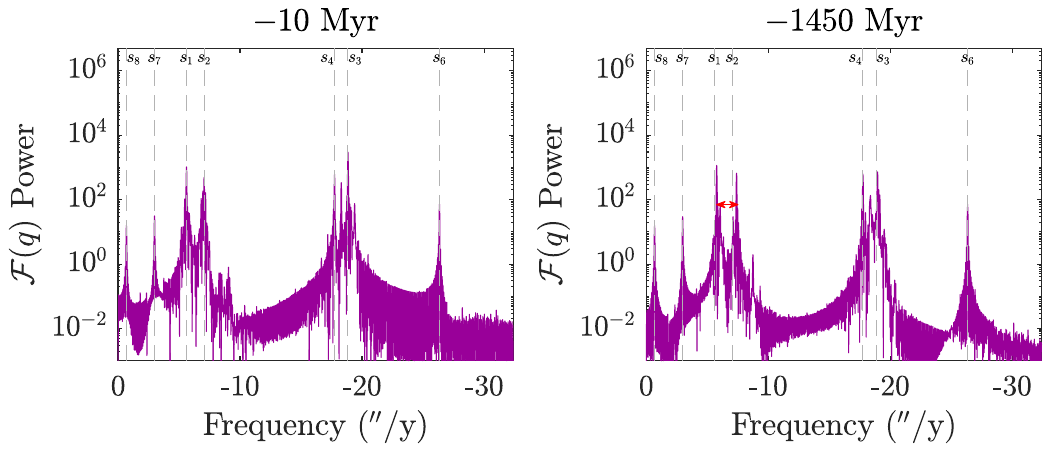}

\vspace*{-50ex}
\caption{\scs
Time series analysis of astronomical solutions.
Top and bottom: $g$- and $s$-spectra determined from our 
solar system integrations using fast Fourier transform (FFT~$= {\cal F}(k,q)$) 
over 20 Myr intervals and Earth's $k$ and $q$ variables (see \meth). 
Frequencies in arcsec~y\pmo\ = ''~y\pmo.
Left: Standard spectra in the recent past centered at
$t = -10$~Myr (nearly identical in all solutions). Right:
spectra of solution R28 centered at $t = -1450$~Myr
(see Fig.~\ref{fig:ecc}b).
The dashed lines indicate frequencies in the recent past (left)
in all panels.
During \sigot-resonance intervals (right), $g_2$ shows
reduced power and generally splits into two peaks (red circles
and inset).
In the recent past (left), $|g_2-g_1| > |s_2-s_1|$, whereas
during \sigot\ resonances (right) $|g_2-g_1| \simeq |s_2-s_1|$ 
(see double arrows).
\label{fig:gsFFT}
}
\end{figure*}

\iftwo \def\vsI{-120ex} \def\vsII{-152ex} \def\vsIII{-093ex}
\else  \def\vsI{-098ex} \def\vsII{-124ex} \def\vsIII{-075ex} 
\fi
\def\hsp{-06ex}
\begin{figure*}[t]
\vspace*{-47ex} \hspace*{\hsp}
\includegraphics[scale=0.90]{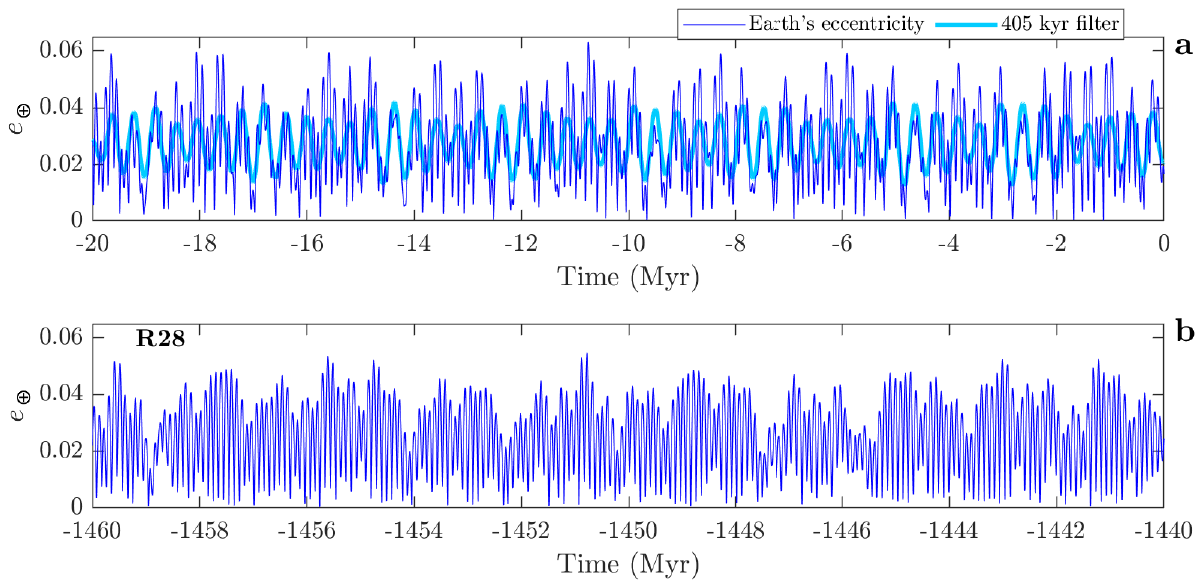}

\vspace*{\vsI}  \hspace*{\hsp}
\includegraphics[scale=0.90]{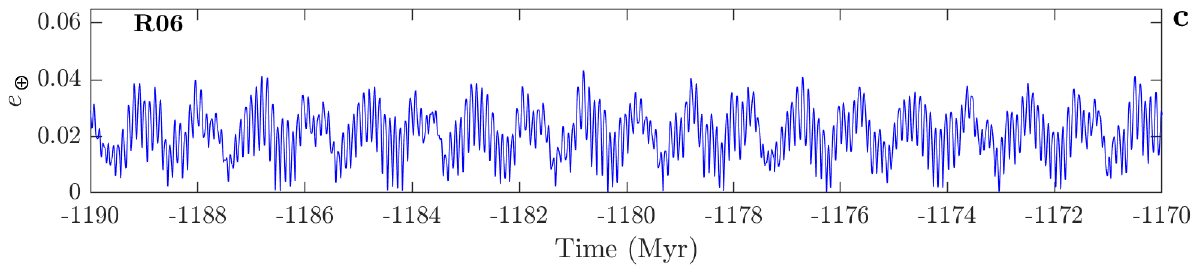}

\vspace*{\vsII} \hspace*{\hsp}
\includegraphics[scale=0.90]{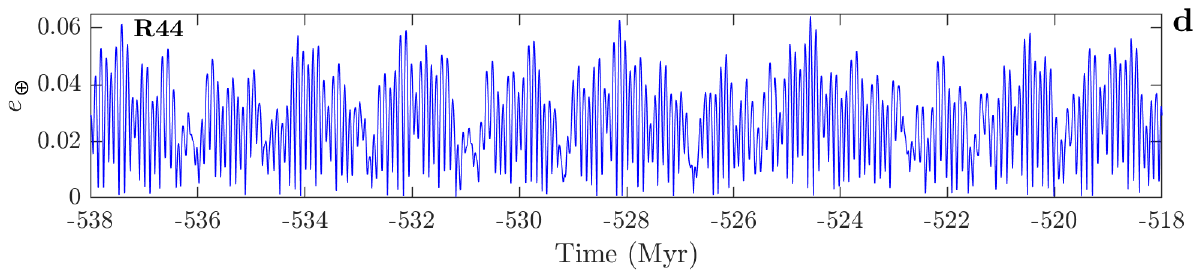}

\vspace*{\vsIII}
\caption{\scs
Earth's orbital eccentricity from ensemble integrations.
(a) Eccentricity pattern in the recent past (last 20 Myr)
with short and long (\sm{100} and \sm{405}~kyr) cycles (nearly
identical in all solutions). Note the strong bundling of four
short cycles into one long cycle (highlighted by 405~kyr filter).
Panels (b, c, d) display examples of solutions during \sigot-resonance intervals.
(b) Solution R28 (Run 28) over a 20-Myr interval centered at
$t = -1450$~Myr. The LEC is virtually absent and the eccentricity
pattern is unrecognizable compared to (a). 
(c) Solution R06 over a 20-Myr interval centered at $t = -1180$~Myr.
In addition to a weak LEC, the maximum eccentricity is reduced to 
\sm{0.04}, which affects the total insolation Earth receives over one 
year (see text).
(d) Solution R44 over a 20-Myr interval centered at $t = -528$~Myr;
the LEC is present but weak relative to the short eccentricity
cycle.
\label{fig:ecc}
}
\end{figure*}

Several observations lend confidence to the robustness 
of our astronomical computations. ($i$) The methods used here and our 
integrator package have been extensively tested and compared
against other studies \citep{zeebe17aj,zeebelourens19,zeebe22aj,
zeebe23aj}. ($ii$) The \sigot\ resonance was recognized previously,
although to our knowledge only by two studies 
\citep{lithwick11,mogavero22} and 
not its effect on \gtf/LEC (see below). 
($iii$) We tested an independent integrator package 
(HNBody \cite{rauch02}, 
see \meth) and found the same dynamical behavior.
($iv$) Re-examination of previous 5-Gyr future integrations
from this group \citep{zeebe15apje} also revealed various 
solutions with \sigot-resonance intervals.
($v$) Total energy and angular momentum errors were 
small throughout our present 3.5-Gyr integrations (relative errors
in test runs: $\lsim$$6 \x 10^{-10}$ and $\lsim$$7 \x 10^{-12}$, 
Fig.~\ref{fig:dEL}) and our numerical timestep sufficiently 
resolves Mercury's perihelion \citep{wisdom15,hernandez22,abbot23} 
(see \meth).

\begin{figure*}[t]
\vspace*{-30ex} \hspace*{-00ex}
\includegraphics[scale=0.8]{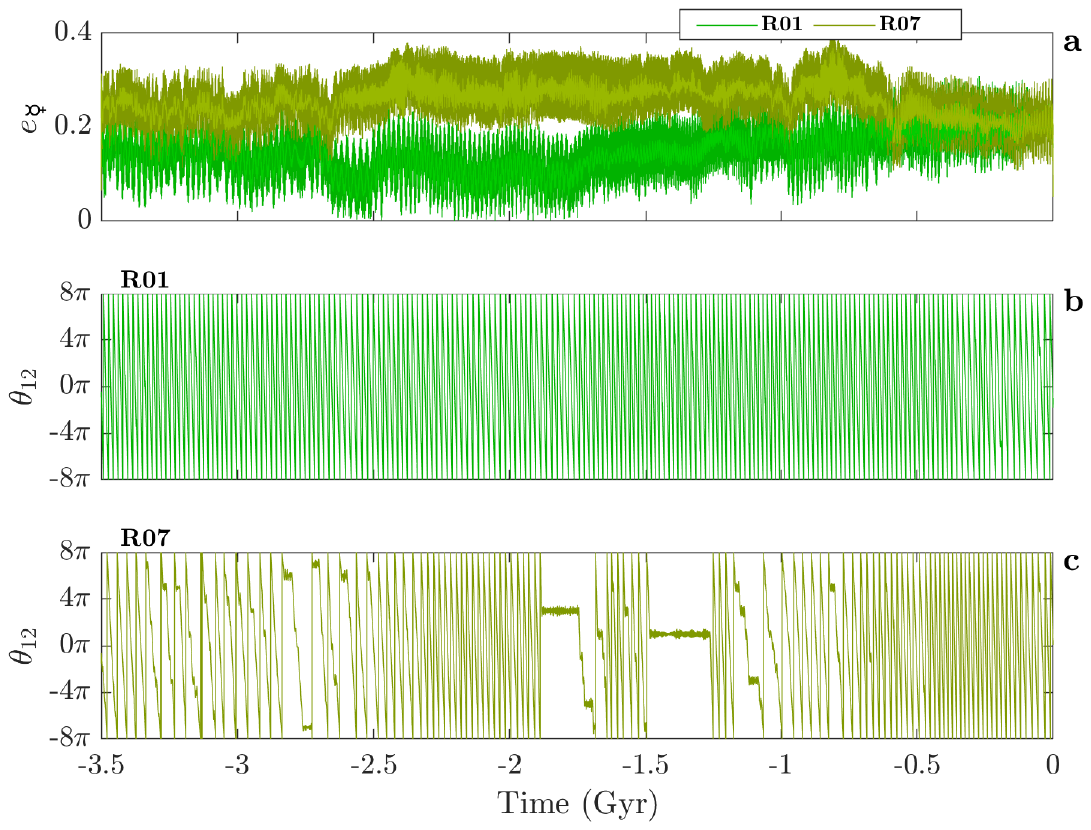}

\vspace*{-30ex}
\caption{\scs
Examples of
Mercury's orbital eccentricity and resonant angle \thtot.
(a) Mercury's orbital eccentricity in solutions R01 and R07.
The lighter foreground colors represent filter magnitudes
($|z^*_M|$, see Section~\ref{sec:rang}).
(b, c) Resonant angle 
$\thtot = (\vpi^*_1 - \vpi^*_2) + (\Om^*_1 - \Om^*_2)$
associated with $\sigot = (g_1 - g_2) + (s_1 - s_2)$
(see Section~\ref{sec:rang} for calculation). 
\thtot\ is intentionally shown over a $16\pi$ range for 
clarity (at $2\pi$ range, lines connect and appear 
as areas/patches).
(b) \thtot\ in R01 circulates throughout;
\sigot-resonance intervals are absent.
(c) \thtot\ in R07 circulates but also librates during \sigot-resonance
intervals (oscillation around a constant value, see e.g., plateau
at about $-1.25$ to $-1.5$~Gyr in (c)).
Solutions exhibiting \thtot-circulation and libration were usually
associated with intervals of slightly elevated eccentricity in 
Mercury's orbit.
\label{fig:theta}
}
\end{figure*}

During \sigot-resonance episodes, Earth's
orbital eccentricity pattern and hence Earth's climate forcing 
spectrum due to eccentricity becomes unrecognizable 
compared to the recent past (Figs.~\ref{fig:EccFFT}
and~\ref{fig:ecc}).
For example, a geologist examining a climate record exhibiting
Milankovi{\'c} cycles within the \sigot\ resonance 
(e.g., Fig.~\ref{fig:ecc}b-d) would fail to identify the rhythm 
as eccentricity cycles, given the currently known pattern 
(Fig.~\ref{fig:ecc}a).
The resonance motifs are so different that (coincidentally)
some frequency 
and amplitude modulation (AM) features (Fig.~\ref{fig:ecc}c)
show more similarities with Mars' orbital inclination in the 
recent past (Fig.~\ref{fig:IMars}) \citep{zeebe22aj} than 
Earth's eccentricity (Fig.~\ref{fig:ecc}a).
The estimated time scale \tauot\ for a possible \sigot-resonance
occurrence ($\tauot =$ temporal distance to the present) is of order
$10^8$ to $10^9$~y. In several solutions, we found reduced 
$g_2$ and \gtf\ power (lower than short eccentricity power),
as well as unusual eccentricity patterns, at $t \ \lsim -500$~Myr
(see e.g., Fig.~\ref{fig:ecc}d) and in one solution at $t \simeq -420$~Myr.
However, we have so far tested only 64 solutions (see \meth), 
hence the youngest possible age of a \sigot-resonance
interval that could be detectable in the geologic record 
is yet unknown. The duration of a \sigot-resonance episode may 
range from a few Myr to tens of millions of years (multiple 
entries/exits often occurring over several 100 Myr,
see Figs.~\ref{fig:ecc} and ~\ref{fig:theta}).

\begin{figure*}[t]
\vspace*{-30ex} \hspace*{+00ex}
\includegraphics[scale=0.8]{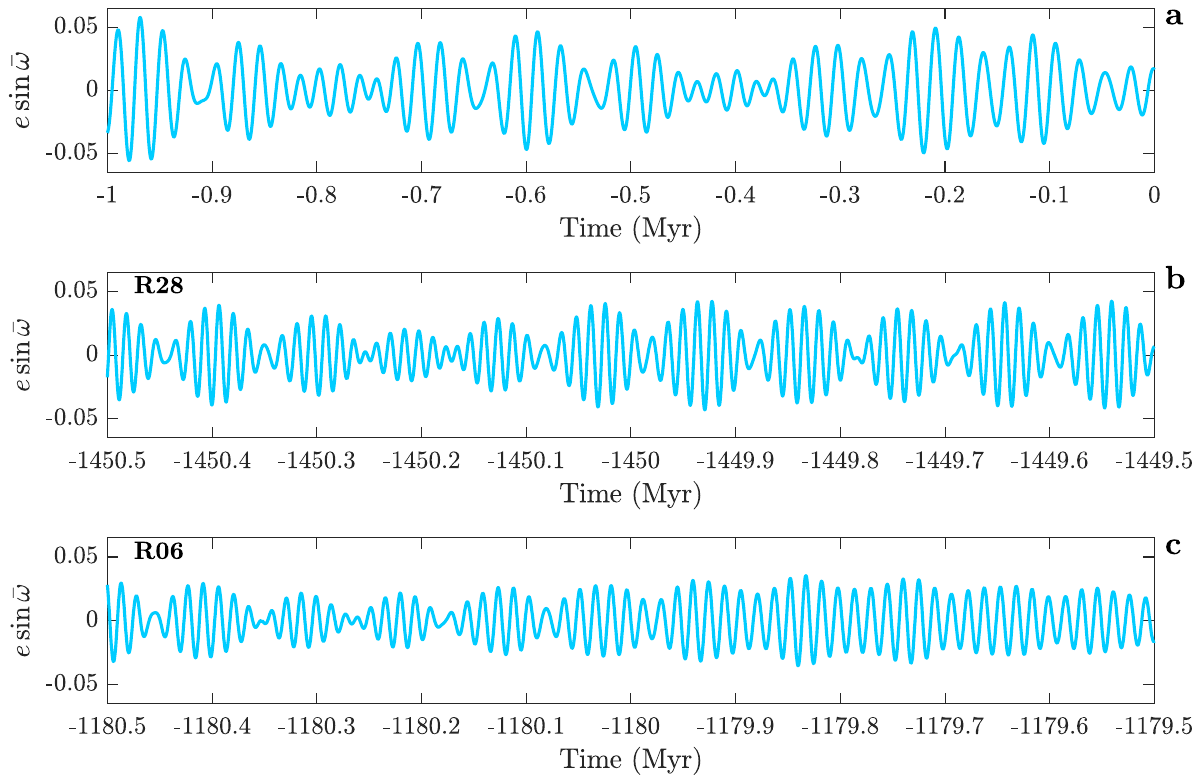}

\vspace*{-35ex}
\caption{\scs
Climatic precession $\bar{p} = e \sin \bar{\omega}$,
where $\bar{\omega}$ is the longitude of perihelion
measured from the moving equinox (for details and code, see 
\citep{zeebe22aj}, \snvurl, \myurl, and Section~\ref{sec:al}. 
(a) $\bar{p}$ in 
the recent past (last 1~Myr), nearly identical in
all solutions.
Panels (b, c) display examples of $\bar{p}$ based
on orbital solutions during \sigot-resonance intervals.
Note the different lunisolar precession frequencies (carrier
frequency) in the past
owing to the Earth-Moon system's evolution.
(b) $\bar{p}$ based on solution R28 over a 1-Myr interval centered at
$t = -1450$~Myr. The amplitude modulation (AM) pattern fundementally
differs from (a).
(c) $\bar{p}$ based on R06 over a 1-Myr interval centered at 
$t = -1180$~Myr. In addition to an altered AM pattern, the total
amplitude is reduced compared to (a).
\label{fig:CP}
}
\end{figure*}

\subsection{Insolation and climatic precession}

\begin{figure*}[t]
\vspace*{-35ex} \hspace*{+00ex}
\includegraphics[scale=0.8]{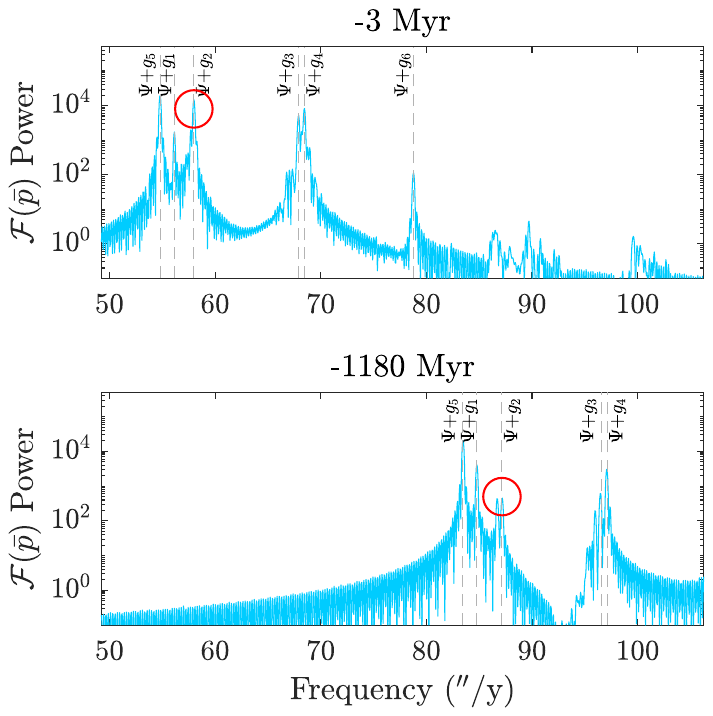}

\vspace*{-40ex}
\caption{\scs
Fast Fourier transform (FFT~$= {\cal F}(\bar{p})$) 
of climatic precession $\bar{p} = e \sin \bar{\omega}$
over 6-Myr intervals.
Frequencies in arcsec~y\pmo\ = ''~y\pmo.
Note different lunisolar precession frequency
$\Psi$ in the past owing to the Earth-Moon system's evolution
(for details and code, see \citep{zeebe22aj}, \snvurl,
\myurl, and Section~\ref{sec:al}).
Top: Standard spectrum in the recent past centered at
$t = -3$~Myr (nearly identical in all solutions). Bottom:
Spectrum based on solution R06 centered at $t = -1180$~Myr.
Note the split and reduced power of $\Psi + g_2$ in bottom vs.\
top panel (red circles).
\label{fig:cpFFT}
}
\end{figure*}

The total mean annual insolation (or energy $W$)
Earth's receives is a function of its orbital eccentricity (\eE):
\beqn
W \propto \left( 1 - \eE^2 \right)^{-\q{1}{2}} \ .
\label{eqn:insol}
\eeqn
In the recent past, $0 \ \lsim \eE \lsim 0.06$, whereas
during \sigot-resonance episodes, $\max\{\eE\}$ may be
as low as \sm{0.04} (Fig.~\ref{fig:ecc}c). Thus, the relative 
variation/difference in $W$ between eccentricity maxima and minima 
$((1 - 0^2)^{-\q{1}{2}} = 1)$
is substantially reduced by the factor:
\beqn 
\left( (1 - 0.06^2)^{-\q{1}{2}} - 1 \right) \ \left/ \ 
\left( (1 - 0.04^2)^{-\q{1}{2}} - 1 \right) \ \right.
\iftwo \nonumber \\ \fi
= 2.25 \ .
\eeqn
Thus, in addition to a weak/absent LEC, both the total variation 
in eccentricity climate forcing and the extreme values are 
diminished on a $10^6$-year time scale during \sigot\ intervals. 
Moreover, the \sigot\ resonance causes major changes in
climatic precession ($\bar{p}$), the primary climate driver on 
the shortest Milankovi{\'c} time scale (\sm{20}~kyr in the recent 
past). The main $\bar{p}$ frequencies are given by $\Psi + g_i$,
where $\Psi$ is the lunisolar precession frequency. The
disruption of $g_2$ (and hence $\Psi + g_2$) causes
major changes in $\bar{p}$'s total amplitude and AM
(see Figs.~\ref{fig:CP} and \ref{fig:cpFFT}).
For example, during \sigot-resonance intervals, the forcing power 
at the $\Psi + g_2$ precession frequency may drop by orders of
magnitudes compared to the recent past (Fig.~\ref{fig:cpFFT}).
The altered forcing in both, eccentricity and climatic 
precession, would scale down the climate response to orbital
forcing, and hence its expression in geological 
sequences, as well as affect threshold behavior for triggering 
orbitally forced climate events (for recent examples such
as the Paleocene-Eocene Thermal Maximum and the Eocene 
hyperthermals, see \cite{zeebelourens19}).

\subsection{\sigot\ resonance}

In a quasi-periodic (non-chaotic) system, the 
fundamental frequencies are constant over time.
In contrast, the solar system's frequencies
change over time owing to dynamical chaos
(Fig.~\ref{fig:g.dM}), priming the system to
enter/exit the \sigot\ resonance over long time 
scales.
Several solar system resonances have been studied
previously, including $(g_1 - g_5) - (s_1 - s_2)$ and 
$2(g_4 - g_3) - (s_4 - s_3)$ \citep{laskar90,sussman92,
batygin15,mameyers17,zeebe22aj,abbot23}. However, to our knowledge 
only two studies recognized \sigot, yet did not investigate
its consequences for \gtf\ and Earth's orbital eccentricity
\citep{lithwick11,mogavero22}.
\sigot\ may be characterized by the resonant angle \thtot\
(see Eq.~(\ref{eqn:tht})),
where $\vpi^*$ and $\Om^*$ are associated with the $g$- and
$s$-modes (analog to longitude of perihelion and ascending 
node, but not of individual planets, see Appendix~\ref{sec:gs}).
Chaos
is often associated with resonant angles that alternate
between circulation and libration (Fig.~\ref{fig:theta}).
Generally, \thtot\ circulated in our solutions without \sigot-resonance 
episodes, whereas \thtot-circulation {\sl and} libration occurred in 
solutions that showed \sigot-resonance episodes and a 
weak/absent LEC. The latter case was usually associated with 
intervals of slightly elevated eccentricity in Mercury's orbit
($0.25 \ \lsim \eM \lsim 0.35$, see Fig.~\ref{fig:theta}).

Importantly, none of our solutions showed high eccentricities
($\eM \ \gsim 0.4$), which could indicate progressing chaotic
behavior or a potential 
destabilization of the inner solar system --- known, separate
dynamical phenomena, most relevant to future chaos,
studied previously \citep{laskar90,sussman92,lithwick11,batygin15,
zeebe15apje,brownrein20,abbot23}. Solutions displaying any 
significant destabilization in the past can of course be excluded 
(incompatible with the solar system's known history).
Furthermore, given that all our
solutions showed at most slightly elevated \eM\ demonstrates that
the system can enter/exit the \sigot\ resonance without major changes 
in, or destabilization of, planetary orbits. Thus, past \sigot-resonance 
episodes and a weak LEC are a possible and likely dynamical 
phenomenon, present in \otp\ of our solutions.

\section{Implications}

We anticipate far-reaching consequences of our findings for
($i$) exploring the effects of secular resonances (particularly
\sigot) on the long-term dynamical evolution, chaos, and planetary
climates in the solar system,
($ii$) understanding and
unraveling Earth's past climate forcing and 
climate change via parameters including 
eccentricity (total insolation) and climatic precession 
(see Eq.~(\ref{eqn:insol}) and
Figs.~\ref{fig:EccFFT}, \ref{fig:ecc}, \ref{fig:CP}, \ref{fig:cpFFT}),
($iii$) reconstructing the solar system's chaotic
dynamics constrained by geologic evidence
\citep{mameyers17,meyers18,zeebelourens19,olsen19},
($iv$) expanding
the evidence for the astronomical theory of climate in yet
understudied parts of Earth's history (e.g., the Precambrian),
($v$) studying effects of deep-time Milankovi{\'c} forcing on 
Earth's environmental and climatic long-term evolution,
and ($vi$) extending the astronomically calibrated geological
time scale \citep{montenari18,zeebelourens22epsl} into deep time.

It appears that the \sigot\ secular resonance and its effect
on solar system dynamics and planetary climates has been understudied 
thus far. To our knowledge, only two studies recognized \sigot, yet 
did not investigate its consequences on, for instance, \gtfL\ and 
Earth's orbital eccentricity \citep{lithwick11,mogavero22}. Several 
of the secular modes (or terms) related to the $g_i$ and $s_i$ (or 
differences between pairs) show multiple,
strong interactions for $i = 1, \ldots 4$. In other words, secular 
resonances usually affect multiple frequency 
pairs. For example, as shown here, the \sigot\ resonance has a major 
impact on ($g_1$$-$$g_2$) and ($s_1$$-$$s_2$), but also on \gtfL. 
Moreover, because there is no simple one-to-one 
relationship between a single mode and a single inner planet 
(the motion is a superposition of all modes), resonances (say 
$\sigma_{ij}$ with $i,j = 1, \ldots 4$, $i \neq j$) affect the entire inner 
solar system. Here, we have only investigated \sigot's effect on
\gtfL\ and Earth's orbital eccentricity. Future work should 
explore whether there are other
important, yet unknown, effects of \sigot\ (and other 
resonances) on the dynamics and planetary climates in the 
inner solar system.

Our results have fundamental implications for, e.g., current astrochronologic 
and cyclostratigraphic practices, which are based 
on the paradigm that the LEC is stable, dominates the eccentricity 
spectrum, and has a period of \sm{405}~kyr. Given our findings that 
the \sigot\ resonance is a common phenomenon (occurring in \otp\
of our solutions), the assumption of a stable 405~kyr cycle in deep
time can no longer be made. Specifically, the possibility of 
an unstable period and weakened LEC amplitude requires
rethinking of currently employed strategies for building 
accurate and high-resolution (`floating' or radio-isotopically anchored)
astrochronologic age models. The presumed 405-kyr ``metronome'' was
particularly important for constructing pre-Cenozoic age models,
where reliable astronomical solutions are absent owing
to solar system chaos.
Deep-time astrochronologies have thus far critically relied on identification 
of the LEC because it is the only Milankovi{\'c} cycle whose period 
has been widely regarded as stable \citep{laskar04Natb,
kent18,spalding18} {\sl and} of sufficiently large amplitude 
to be typically expressed in sedimentary sequences.
Our results indicate that, prior to several hundred Myrs
in the past, the LEC may have become unstable over 
multi-million year intervals. Notably, on these time scales the 
periods of other critical Milankovi{\'c} parameters (climatic 
precession and obliquity) are also more uncertain due to changes 
in the Earth-Moon system's tidal evolution.

Does the presently explored geologic record provide
examples consistent with our astronomical calculations? We note 
here a recently discovered section in the \sm{2.46}~Ga Joffre 
Member of the Brockman Iron Formation (Joffre Falls, Western Australia), 
in which a dominant short eccentricity cycle was reported
(\sm{100}~kyr), compared to a relatively weak expression at the scale 
of the interpreted LEC \citep{lantink22} (see 
Fig.~\ref{fig:Joffre}).
The interpreted eccentricity modulation pattern in the 
Joffre Falls section differs from typical Cenozoic precession-eccentricity
dominated records, which often display a strong 1:4 hierarchy of 
long vs.\ short eccentricity cycles. 
One first-order interpretation for the unusual bundling 
pattern is a complex nonlinear response of the 
paleoclimate and/or sedimentary system to orbital forcing, 
which is still poorly understood for the ancient deposits. Yet, 
given our findings, a fundamentally different 
pattern of Earth's orbital eccentricity variations 
(i.e., a weakened LEC) at the time of deposition provides an
alternative explanation (see Fig.~\ref{fig:Joffre}).
Further investigations are required to confirm past
\sigot-resonance episodes in geologic sequences.
We propose that future exploration of high-quality and rhythmic 
sediment successions (especially of Precambrian age) will be 
critical in constraining the LEC's past stability and hence 
the history of the solar system’s chaotic evolution.

Generally, the possibility of an unstable/weak LEC argues strongly for 
internal consistency checks and tests of eccentricity-related 
cycles interpreted in stratigraphic sequences at multiple levels. 
For example, future studies need to include consistency checks 
between the period of short eccentricity and associated 
$g$-frequencies and the period of the interpreted $(g_4-g_3)$ cycle 
and/or other very long period eccentricity modulations. At a 
more advanced level, the internal consistency of all $g$- and 
$s$-frequencies 
that can be extracted from the sequence need to be examined,
for which algorithms are already available
\citep{meyers18,olsen19}. Furthermore, the uncertainty in LEC 
stability substantially increases the ambiguity in 
interpreting cycle ratios
of the eccentricity-precession forcing. Hence,
independent sedimentation rate checks (preferably from 
accurate radiometric ages) will become inevitable to verify
deep-time
Milankovi{\'c} interpretations based on observed stratigraphic
cycle hierarchy. Moreover,
when significant obliquity signals are present,
eccentricity-related cycle pattern will be more difficult to 
distinguish from those expected for obliquity.


\begin{acknowledgments}
{\bf Acknowledgments.} 
This research was supported by Heising-Simons Foundation Grants 
\#2021-2800 and \#2021-2797 (R.E.Z. and M.L.L.) and U.S. NSF grants 
OCE20-01022, OCE20-34660 to R.E.Z. We thank the reviewer 
for suggestions, which improved the manuscript. \\
\end{acknowledgments}


\software{
          \orb, \giturl; on Zenodo: \zenurl,
          \cite{orbitn23}.
          }

\appendix

\renewcommand\thefigure{\thesection.\arabic{figure}}    
\renewcommand{\baselinestretch}{0.7}
\small

\section{$g$- and $s$-modes \label{sec:gs}}
\setcounter{figure}{0}    

\begin{figure}[h]
\vspace*{-15ex} \hspace*{-03ex}
\includegraphics[scale=0.45]{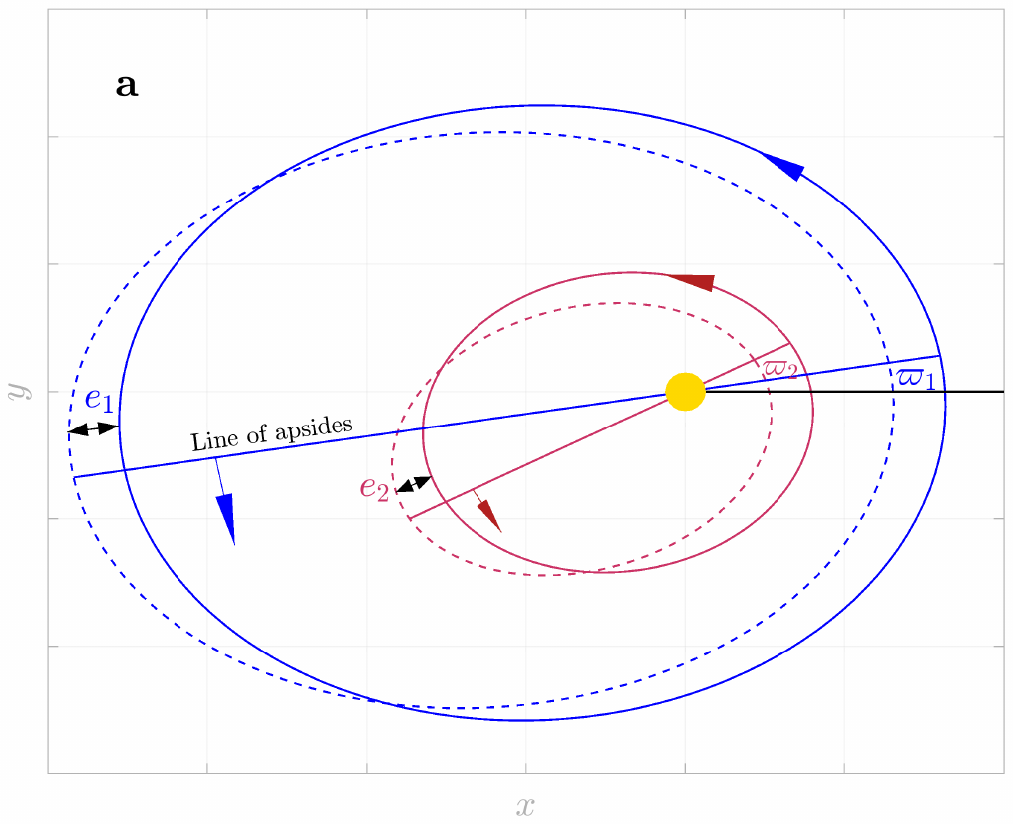}
\vspace*{-00ex} \hspace*{-05ex}
\includegraphics[scale=0.45]{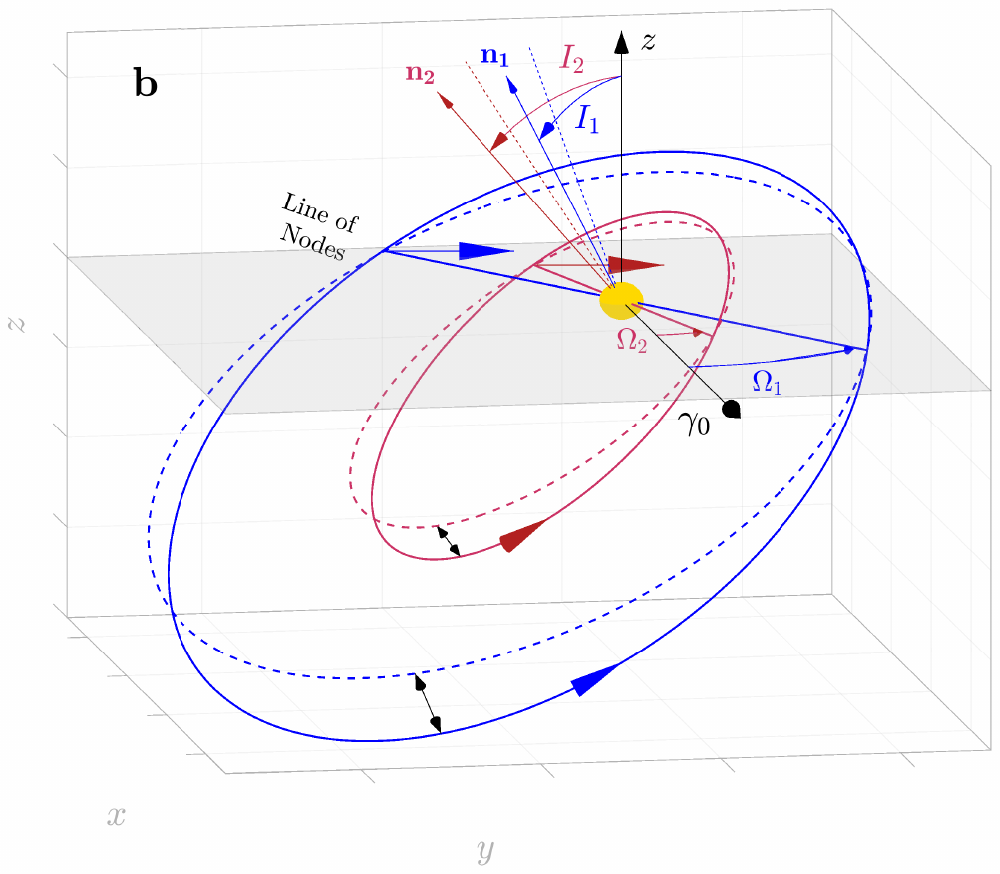}

\vspace*{-20ex}
\caption{\scs
Schematic illustration of (a) $g$- and (b) $s$-modes (see text). 
\label{fig:gsill}
}
\end{figure}

The $g$- and $s$-modes and their interaction is central
for the secular resonances discussed in this study 
(for illustration, see Fig.~\ref{fig:gsill}).
Note that there is generally
no simple one-to-one relationship between a single mode 
and a single planet, though some outer planets may dominate a single 
mode. $e$, $I$, $\vpi$, and $\Om$ are eccentricity, 
inclination, longitude of perihelion, and longitude of ascending 
node, respectively. 
$g$-modes are related to $e$ and $\vpi$;
$e$ usually varies between some extreme values (black double arrows) and
$\vpi$ characterizes the apsidal precession; $\vpi$ may librate (oscillate) 
or circulate for solar system orbits. The planetary $g_i$'s
are positive, hence for circulating $\vpi$, the time-averaged apsidal 
precession is prograde (i.e., in the same direction as the orbital 
motion, see large colored arrows). 
The $s$ modes are related to $I$ and $\Om$; $I$ usually varies
between some extreme values (black double arrows) and 
$\Om$ characterizes the nodal precession; $\Om$ may librate 
or circulate. The planetary $s_i$'s are negative, 
hence for circulating $\Om$, the time-averaged nodal precession is 
retrograde (i.e., in the opposite direction as the orbital motion, 
see large colored arrows). Given conservation
of total angular momentum ($\vb{L}$), there exists an invariable 
plane perpendicular to $\vb{L}$, which is fixed in space. It
follows that one of the $s$ frequencies is zero ($s_5$, see 
main text). 

\section{Period of \gtfL}
\setcounter{figure}{0}    

\begin{figure}[h]
\vspace*{-50ex} \hspace*{-00ex}
\includegraphics[scale=0.8]{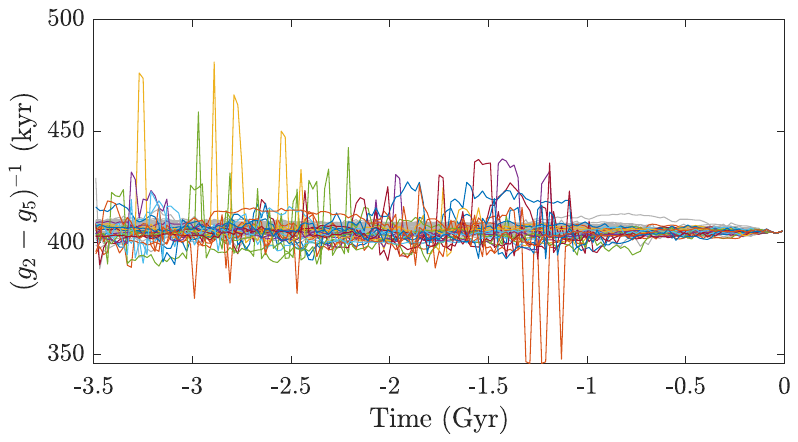}

\vspace*{-50ex}
\caption{\scs
Period of $\gtf = (g_2-g_5)$ from our ensemble integrations.
Solutions including \sigot-resonance intervals (\otp)
are highlighted in color, the remaining solutions are displayed
in gray.
\label{fig:T25}
}
\end{figure}

The secular frequency $g_2$ shows large and rapid shifts (spikes,
see Fig.~\figgdM) at specific times when the 
spectral $g_2$ peak splits into two peaks at significantly reduced
power during \sigot-resonance episodes (see Fig.~\figgsFFT).
Alternating maximum power
between the two peaks then causes the spikes in $g_2$ and 
hence in \gtf\ (Fig.~\ref{fig:T25}). 
As a result, \gtf\ is unstable and weak/absent 
during \sigot-resonance intervals. 
Note that for, e.g., geological applications, the weak/absent
LEC is crucial, not the actual value of the \ptf\ shift, which is 
immaterial because it would be unidentifiable in a
stratigraphic record owing to the low \gtf\ power.

\section{Mars' inclination}
\setcounter{figure}{0}    

\begin{figure}[h]
\vspace*{-41ex} \hspace*{10ex}
\includegraphics[scale=0.7]{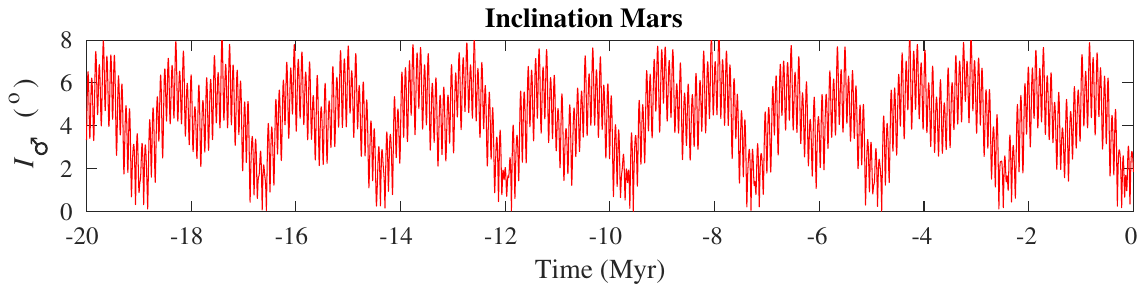}

\vspace*{-45ex}
\caption{\scs
Mars' inclination over the past 20 Myr (nearly identical in all 
solutions, reference frame: ECLIPJ2000).
\label{fig:IMars}
}
\end{figure}

The \sigot-resonance motifs are fundamentally different 
from the recent past (Fig.~\ref{fig:ecc}), some of which
(coincidentally) 
show more similarities with Mars' orbital inclination
in the recent past (Fig.~\ref{fig:IMars}) than Earth's 
eccentricity.

\iftwo 
\else
\clearpage
\fi

\section{Section at Joffre Falls, Western Australia}
\setcounter{figure}{0}    

\begin{figure}[h]
\vspace*{-00ex} \hspace*{+20ex}
\includegraphics[scale=0.4]{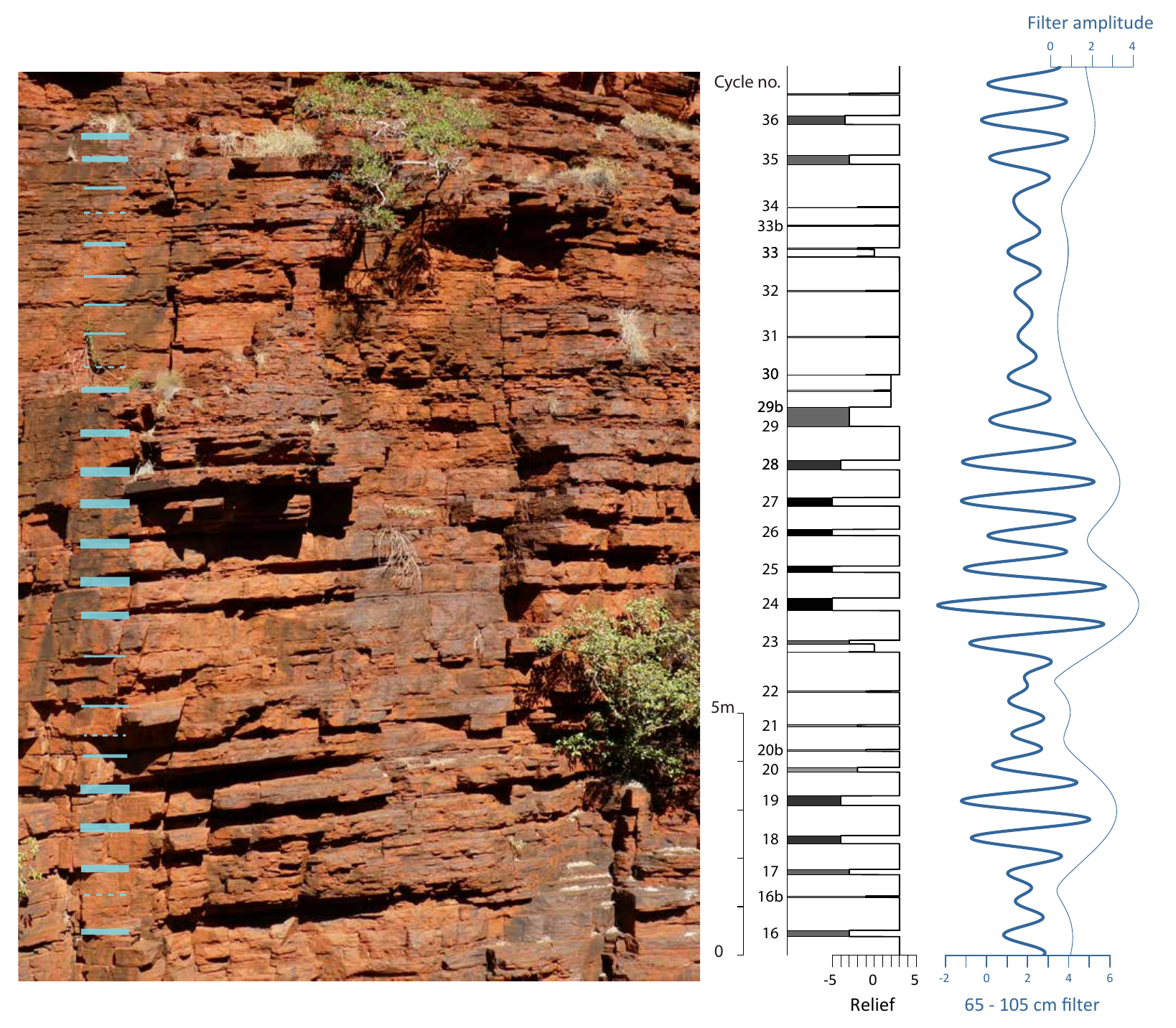}
\vspace*{-02ex}
\caption{\scs
Section of the \sm{2.46}~Ga Joffre Member exposed at Joffre Falls 
(Joffre Gorge, Karijini National Park, Western Australia). 
Photo credit: Frits Hilgen.
\label{fig:Joffre}
}
\end{figure}

In the recently discovered Joffre Falls section (\sm{2.46}~Ga, Western 
Australia), a dominant short eccentricity cycle was reported (\sm{100}~kyr)
and a relatively weak expression of the LEC \citep{lantink22} 
(see Fig.~\ref{fig:Joffre}). The regular 
medium-scale (\sm{85}~cm) alternations of thicker units of banded 
iron formation and thinner, softer interval of a more shaley lithology
(Fig.~\ref{fig:Joffre}, left) have been interpreted as the expression of 
short eccentricity \citep{lantink22}.
Horizons highlighted in blue (Fig.~\ref{fig:Joffre}, left) correspond 
to cycle numbers in the original log \citep{lantink22} shown on the right. 
Note the larger-scale modulations in the thickness and relief 
of the shaley beds (degree of weathering) forming two distinctive darker
'bundles' defined by cycles 17-19 and 23-28. This pattern deviates from 
an expected strong \sm{1:4} bundling pattern or ratio between the 
medium- and large-scale cycles in case of a strong and stable LEC. 
(However, note the \sm{1:4} ratio visible in the filter amplitude of 
the bandpass filtered cyclicity on the right.) Higher up in the 
stratigraphy, the shaley layers become weaker overall and thus any 
larger-scale modulations are more difficult to recognize. Nevertheless, 
we count at least 6 medium-scale cycles until the next more distinctive 
shaley interval, i.e., again no clear 1:4 ratio as would
be expected in case of a strong and stable LEC.

\clearpage


\bibliographystyle{aasjournal}





\end{document}